\documentclass[journal,transmag]{IEEEtran}

\usepackage{graphicx,subfig,color,bm}
\usepackage{amsmath,amsfonts,amssymb,amscd}

\usepackage{booktabs}

\newcommand{\B}{\mathbf{H}}

\begin{document}
\title{Space-Time Galerkin Projection of Electro-Magnetic Fields}%
\author{
\IEEEauthorblockN{Zifu Wang\IEEEauthorrefmark{1}, Thomas Henneron\IEEEauthorrefmark{2}, Heath Hofmann\IEEEauthorrefmark{1}}\\
    	{\IEEEauthorrefmark{1} University of Michigan, EECS, Ann Arbor, MI 48109 USA}\\
	{\IEEEauthorrefmark{2} Universit\'e Lille 1, L2EP, Villeneuve d'ASCQ, Nord 59655 France}\\
    	{zifu.wang@outlook.com}
}
\markboth{Compumag 2015}
{{Wang \MakeLowercase{\textit{et al.}}: Space-Time Galerkin Projection of Electro-Magnetic Fields}}



\maketitle
\begin{abstract}
Spatial Galerkin projection transfers fields between different meshes. 
In the area of finite element analysis of electromagnetic fields, it provides great convenience for remeshing, multi-physics, domain decomposition methods, etc. 
In this paper, a space-time Galerkin projection is developed in order to transfer fields between different spatial and temporal discretization bases.  
\end{abstract}
\begin{IEEEkeywords}
Finite element methods, Galerkin method, Modeling, Field projection.
\end{IEEEkeywords}
\section{Introduction}
The coupling of two finite element analyses (FEA) arises in various applications. 
In the case of multi-physics modeling, for instance, such coupling allows the use of different computation packages for different subproblems (e.g. electromagnetic, thermal or mechanical). 
Since FEA packages tend to use an optimal discretization basis (spatial \& temporal) for specific problems, the communication between different analyses becomes a challenge to be tackled. 

Although interpolation enjoys the advantages of simplicity and high computational efficiency, using it as the communication tool suffers from low precision of transferred field distribution.
Galerkin projection draws greater interest for high-precision coupling, in particular for spatial mesh-to-mesh field transfer \cite{Geuzaine99,Belgacem2001,Farrell2011,transmag2013}.

In this paper, we develop a mesh-to-mesh Galerkin projection method which also transfers fields between different time-discretization bases. 
The proposed method is applied to the magneto-mechanical modeling of an electric machine.
 
\section{Space-Time Projection}
For illustration, in the following, we consider the projection of magnetic field $\B$. 
The proposed projection methods can, however, be generalized to all electromagnetic fields. 

We denote by $\B_{s}$ the magnetic field given on the {\it source} mesh and by $\B_{t}$ the field to be calculated on the {\it target} mesh. 
They are also discretized using different time steps. Here $\B_s$ is supposed to be calculated using $\B$-conforming formulation (e.g. the formulation based on scalar potential $\Omega$). The spatial and temporal variation $\B_s(x,t)$ is thus known (using interpolation) on the domain $D$ during the time $T$.

We define the space-time error norm between $\B_t$ and $\B_s$:
\begin{equation} 
\varepsilon_{\B} =\int_{T} \int_{D} \frac{\mu}{2}\lVert\B_t-\B_s\rVert^2	\label{eq_error}
\end{equation}
where $\mu$ is the magnetic permeability and varies depending on materials. 
It is used here to energetically weight the error norm \cite{transmag2014a,transmag2014b}.
For simplicity, $\mu$ is linear and time-independent in this paper.

Since $\B_t\in\mathbf{H}(\mathbf{curl})$, edge elements offer the most suitable spatial discretization base. 
In terms of temporal discretization, we use a linear interpolation function.
Thus $\B_t$ writes:
\begin{equation} 
\B_t(x,t)=\sum_{i=1..M,\ j=1..N}\mathbf{w}_i^e(x) w_j^t(t)	X_{ij}	\ ,	\label{eq_bt}
\end{equation}
with $M$ the number of edges of the target mesh, $N$ the number of time steps of the target temporal base, $\mathbf{w}_i^e(x)$ the (spatial) interpolation function associated to the {\it i}-th edge, $w_j^t(t)$ the (temporal) interpolation function associated to the {\it j}-th time step and $X_{ij}$ the degrees of freedom. 

The objective of the Galerkin projection is to find the target field $\B_t(x,t)$, which minimizes the defined space-time error norm \eqref{eq_error}.
Thus the derivatives with respect to all degrees of freedom are zero: $\forall i\in\left\{1..M \right\}, \forall j\in\left\{1..N \right\}$
\begin{eqnarray}
\frac{\partial  }{\partial X_{ij} } \varepsilon_{\B}=0\\
\int_{T} \int_{D} \frac{\mu}{2}\frac{\partial  }{\partial X_{ij}} (\lVert\B_t\rVert^2+\lVert\B_s\rVert^2-2\B_t\cdot\B_s)=0\\
\int_{T} \int_{D} {\mu}(\mathbf{w}_i^e w_j^t\cdot\B_t-\mathbf{w}_i^e w_j^t\cdot\B_s)=0\\
\int_{T} \int_{D} {\mu}(\mathbf{w}_i^e w_j^t\cdot\sum_{k,l}\mathbf{w}_k^e w_j^t	X_{kl}-\mathbf{w}_i^ew_j^t\cdot\B_s)=0
\end{eqnarray}
 
It can be written as matrix equation:
\begin{equation} 
[A][X][B]=[C]\ ,\label{eq_ls}
\end{equation}
where $[X]$ is the $M\times N$ matrix of degrees of freedom, $A_{ij}= \int_{D} \mu\mathbf{w}_i^e\cdot\mathbf{w}_j^e$ is the $M\times M$ matrix of inner products of edge functions, $B_{ij}= \int_{T}w_i^tw_j^t$ is the $N\times N$ matrix of inner products of temporal interpolation functions, and $C_{ij}=\int_{T} \int_{D} \mu\mathbf{w}_i^e w_j^t\cdot\B_s$ is the $M\times N$ matrix of source terms.

Using the Kronecker product of the matrices $[B]^T$ and $[A]$, we can transform the matrix equation \eqref{eq_ls} into a linear system:
\begin{equation} 
([B]^T\otimes [A])[vecX] =[vecC]\ ,\label{eq_ls2}
\end{equation}
where $[vecX]$ and $[vecC]$ are the ordered stock of columns of matrix $[A]$ and $[C]$ respectively.  
This linear system is then solved using iterative methods (e.g. the conjugate gradient method).
\section{Application}

Let us consider the magnetic-mechanical-coupled modeling of an electric machine. 
For electric machines rotating at low speeds (e.g., less than 5000rpm), the main source of vibrations and acoustic noise is usually electromagnetic fields. 

We can chain magnetic and mechanical FEA in order to simulate such vibrations.
In the magnetic study, we rotate the rotor to different positions and calculate the distribution of magnetic fields.
In the mechanical study, we calculate vibrations of the stator and the housing structure using the resulting magnetic forces.

\begin{figure}[b]
\begin{center}
\subfloat[]{
\includegraphics[width=4.5cm]{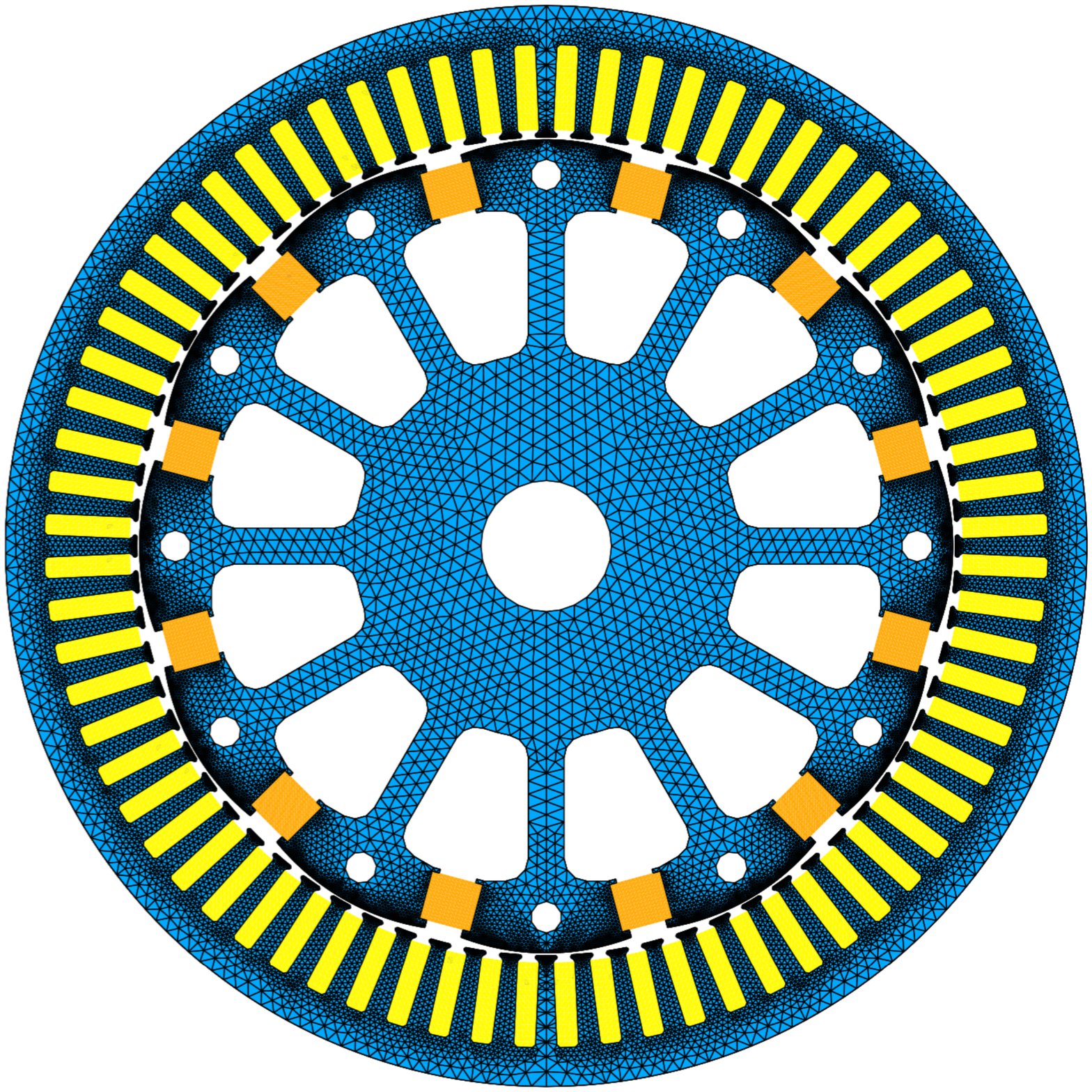}
}
\subfloat[]{
\includegraphics[width=4.5cm]{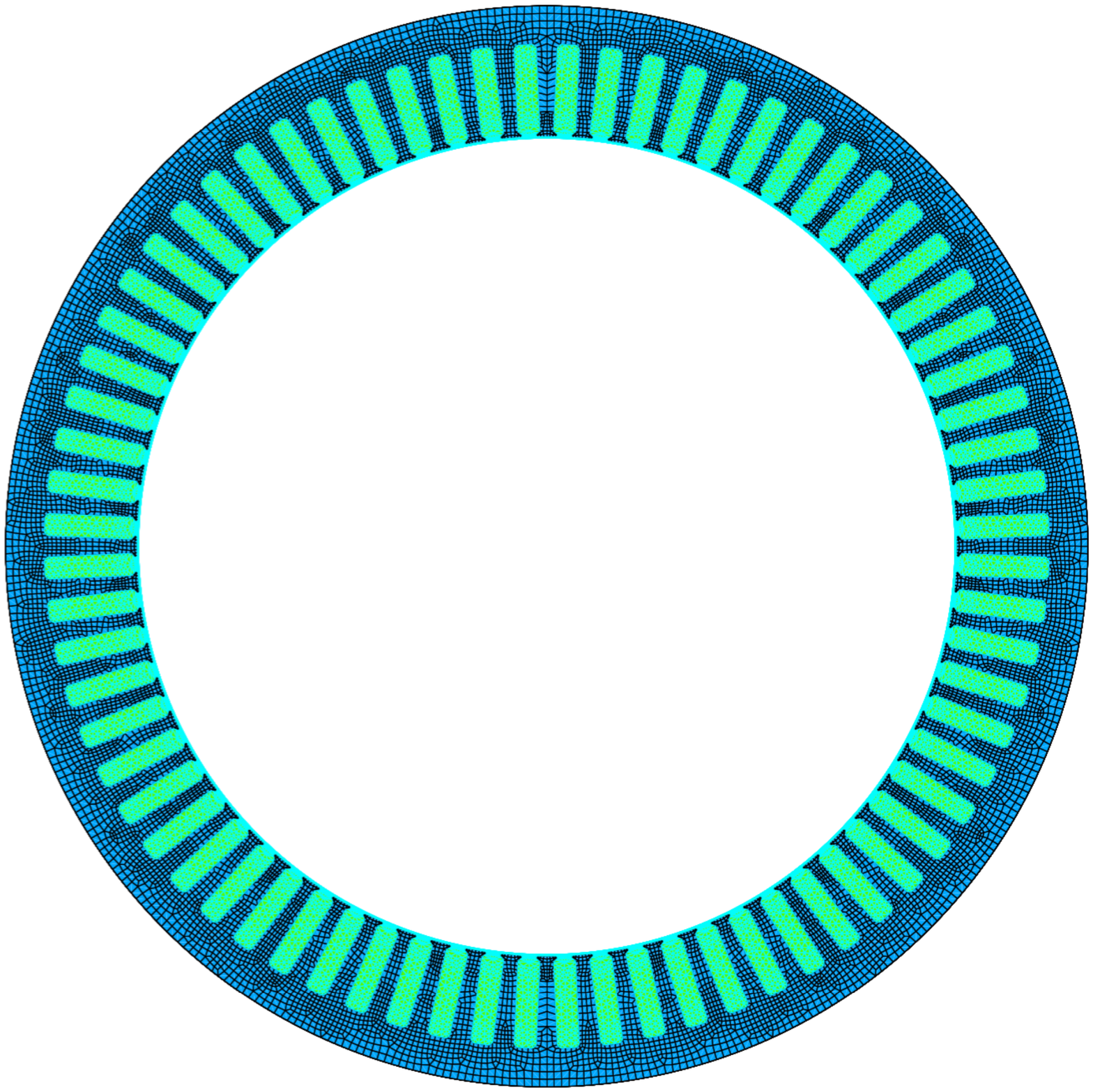}
}
\caption{Meshes used in the magnetic and mechanical models. (a) magnetic mesh: 71k prisms $\times$ 1 layer. (b) mechanical mesh: 12k hexahedrons $\times$ 4 layers.\label{fig_meshes}}
\par\end{center}
\end{figure}

In this case, a reasonable mechanical mesh is often different from the magnetic one, given that the studied geometry is different.
Here the meshes used are illustrated in Fig. \ref{fig_meshes}.
For simplicity, only the stator is considered in the mechanical model.

The projection method presented in the previous section, is thus used to transfer information between the magnetic and mechanical studies.
In particular, the magnetic problem is solved using an $\B$-conforming formation, and the obtained magnetic field $\B$ is projected onto the mechanical mesh. 

Regarding temporal discretization, for the magnetic problem, a fixed-step technique is used to take into account the movement of the rotor. 828 steps are used in the magnetic computation.. 

\begin{figure}[htbp]
\begin{centering}
\includegraphics[width=7cm,bb=0bp 0bp 1125bp 978bp,clip]{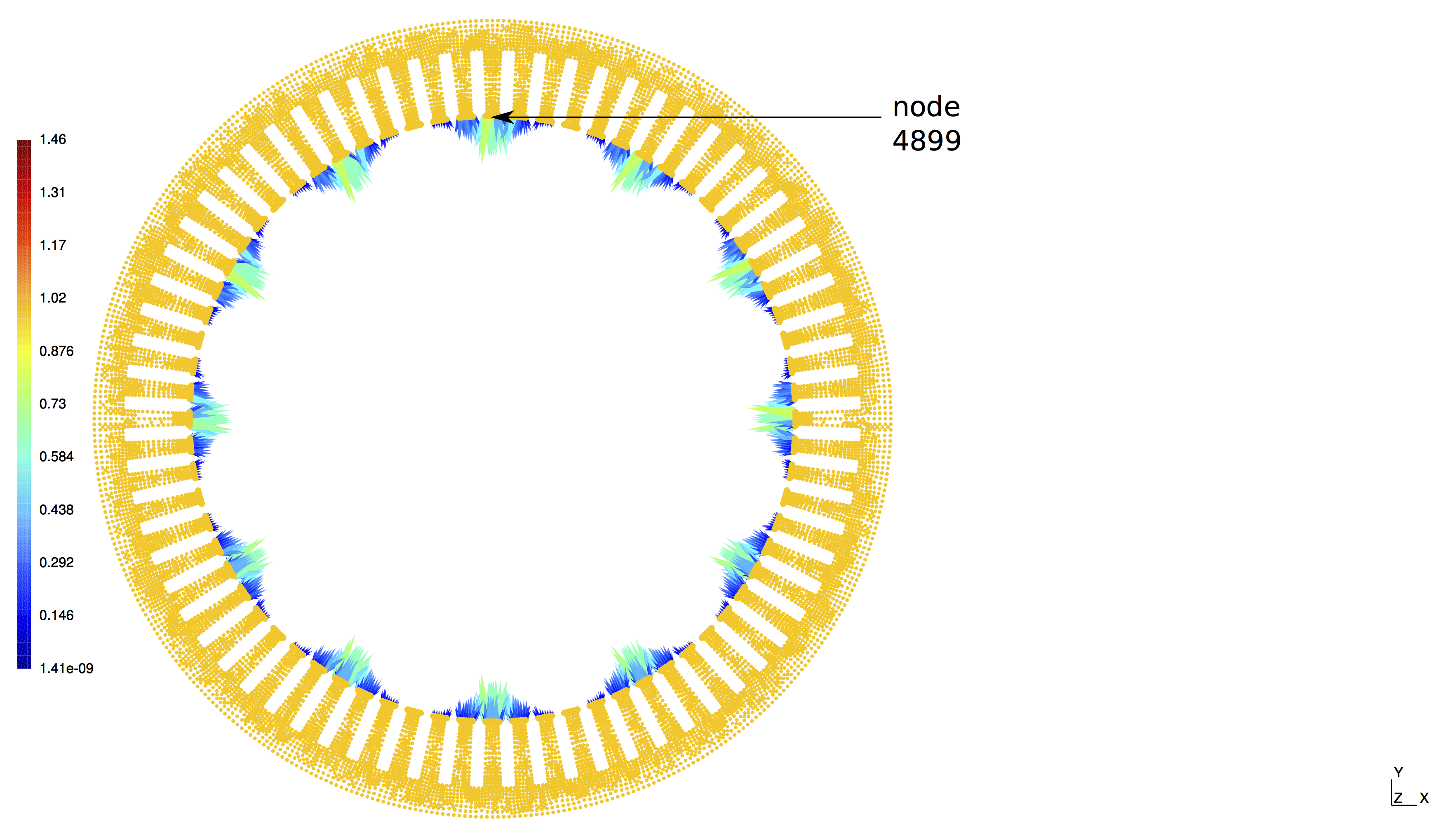} 
\caption{Obtained nodal forces ($N$) on the stator teeth (mechanical mesh), calculated from the space-time projected $\B$ field at $t=0s$. \label{fig_forces}}
\par\end{centering}
\end{figure}

The obtained magnetic field $\B$ is then projected to the mechanical mesh (with edge elements) using \eqref{eq_ls2}. 
Also, the total number of time steps is doubled to 1656 in order to investigate high-frequency components of forces. 
Once the magnetic field $\B$ is projected to the mechanical mesh, the magnetic forces can be obtained by applying the virtual work principle. 
Fig.~\ref{fig_forces} shows result at the first time step ($t=0s$).

\begin{figure}[htbp]
\begin{centering}
\includegraphics[width=8cm]{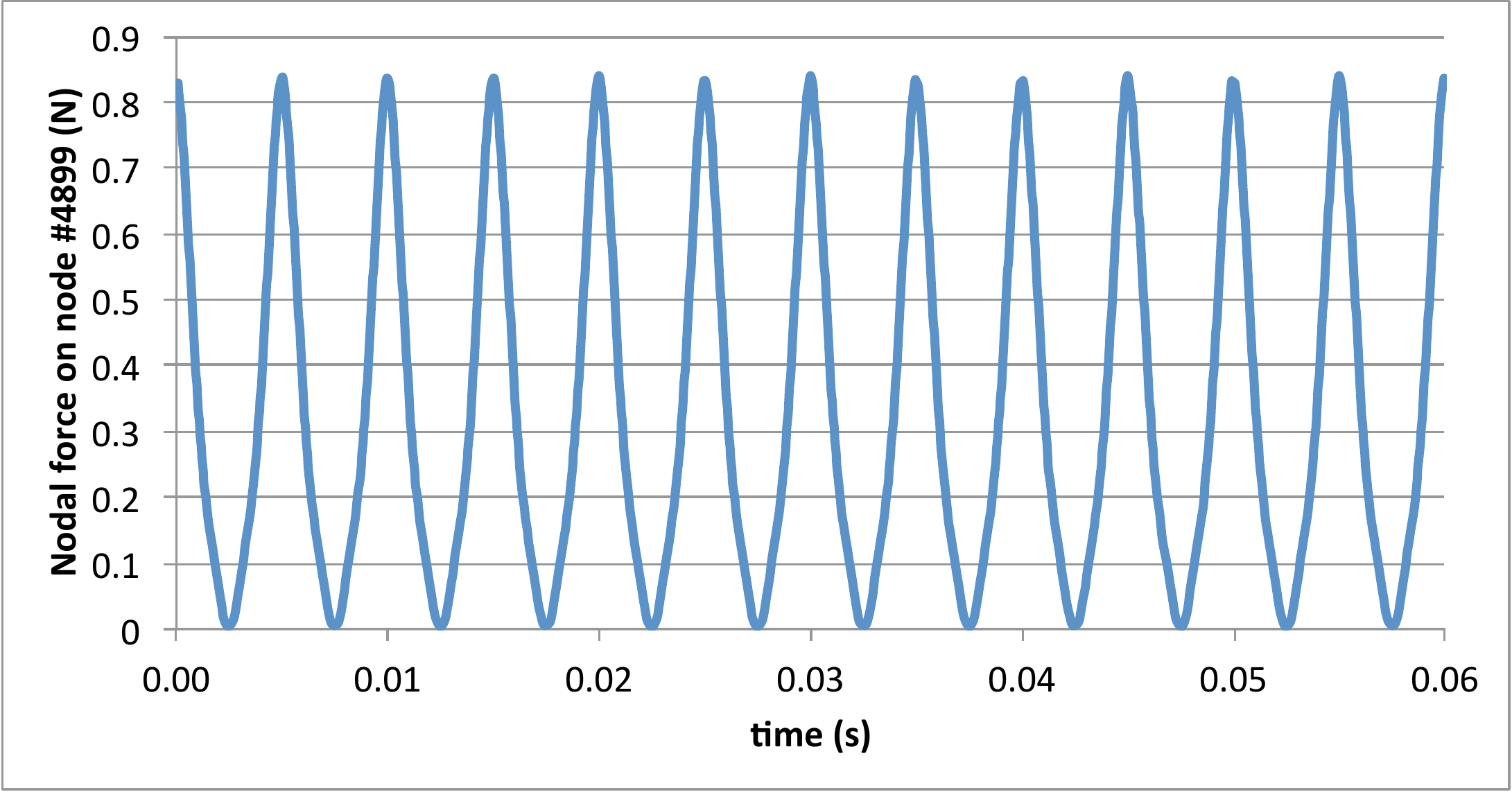} 
\caption{Temporal evaluation of a nodal force ($N$) at the center of a stator tooth (mechanical mesh, node \# 4899), calculated from the space-time projected $\B$ field. \label{fig_temporal_forces}}
\par\end{centering}
\end{figure}

Taking one node on the stator tooth as example, Fig.~\ref{fig_temporal_forces} presents the temporal evaluation of nodal force during one revolution of the rotor. 
Since the studied motor has 12 poles, the nodal force has 12 peaks in one period. 
 
\section{Conclusion}

In this paper, we extend a spatial Galerkin projection into the time domain. 
The proposed projection methods allow the transfer of field distribution between different meshes, and also different temporal discretization bases.

\bibliographystyle{IEEEtran}
\bibliography{biblio}

\end{document}